# In the Name of the Name: RDF Literals, ER Attributes, and the Potential to Rethink the Structures and Visualizations of Catalogs

Manolis Peponakis

## ABSTRACT

*The aim of this study is to contribute to the field of machine-processable bibliographic data that is suitable for the Semantic Web. We examine the Entity Relationship (ER) model, which has been selected by IFLA as a "conceptual framework" in order to model the FR family (FRBR, FRAD, and RDA), and the problems ER causes as we move towards the Semantic Web. Subsequently, while maintaining the semantics of the aforementioned standards but rejecting the ER as a conceptual framework for bibliographic data, this paper builds on the RDF (Resource Description Framework) potential and documents how both the RDF and Linked Data's rationale can affect the way we model bibliographic data.*

*In this way, a new approach to bibliographic data emerges where the distinction between description and authorities is obsolete. Instead, the integration of the authorities with descriptive information becomes fundamental so that a network of correlations can be established between the entities and the names by which the entities are known. Naming is a vital issue for human cultures because names are not random sequences of characters or sounds that stand just as identifiers for the entities—they also have socio-cultural meanings and interpretations. Thus, instead of describing indivisible resources, we could describe entities that appear in a variety of names on various resources. In this study, a method is proposed to connect the names with the entities they represent and, in this way, to document the provenance of these names by connecting specific resources with specific names.*

## INTRODUCTION

The basic aim of this study is to contribute to the field of machine-processable bibliographic data. As to what constitutes "machine processable" we concur with the clarification of Antoniou and van Harmelen, who state, "In the literature the term machine-understandable is used quite often. We believe it is the wrong word because it gives the wrong impression. It is not necessary for intelligent agents to understand information; it is sufficient for them to process information effectively, which sometimes causes people to think the machine really understands."[1] Also, in the bibliography used, the term "computationally processable" is used as a synonym to "machine-processable."

**Manolis Peponakis** (epepo@ekt.gr) is an information scientist at the National Documentation Centre, National Hellenic Research Foundation, Athens, Greece.

With regard to machine-processable bibliographic data, we have taken into consideration both the practice and theory of Library and Information Science (LIS) and Computer Science. From LIS we have chosen the Functional Requirements for Bibliographic Records (FRBR) and the Functional Requirements for Authority Data (FRAD) while making comparisons with the Resource Description and Access (RDA) standard. From the Computer Science domain we have chosen the Resource Description Framework (RDF) as a basic mechanism for the Semantic Web. We examine the Entity Relationship (ER) model (selected from IFLA as a "conceptual framework" for the development of FRBR), [2] as well as the potential problems that may arise as we move towards the Semantic Web. Having rejected the ER model as a conceptual framework for bibliographic data, we have built on the potential of RDF and document how its rationale affects the modeling process.

In the context of the Semantic Web and Uniform Resource Identifiers (URIs), the identification process has been transformed. For this reason we have performed an analysis of appellations and names as identifiers and also explored how we could move on from an era where controlled names play the role of identifiers to one of the URI dominion: "While it is self-evident that labels and comments are important for constructing and using ontologies by humans, the OWL standard does not pay much attention to them. The standard focuses on the syntax, structure and reasoning capabilities. . . . If the Semantic Web is to be queried by humans, there will be no other way than dealing with the ambiguousness of human language."[3]

It is essential to build on the "library's signature service, its catalog,"[4] and use it to provide added-value services. But to get there, first there has to be "a shift in perspective, from locked-up databases of records to open data shared on the Web."[5] This requires a transition from descriptions aimed at human readers to descriptions that put the emphasis on computational processes to escape the rationale of records being a condensed description in textual form and move towards more flexible and fruitful representations and visualizations.

## BACKGROUND

### FRBR and RDA

The FR family has been growing for more than a decade. The first member of the family was the Functional Requirements for Bibliographic Records (FRBR),[6] the first version of which was published towards the end of the last century. Subsequently, IFLA decided to extend the model in order to cover authorities. During this process, the task of modeling the names was separated from the task of modeling the subjects. Thus two new members were added to the family; the "Functional Requirements for Authority Data: A Conceptual Model" (FRAD) and the "Functional Requirements for Subject Authority Data (FRSAD)." [7,8] At the same period of time, the "Resource Description and Access" (RDA) standard was established as a set of cataloging rules to replace the AACR standard. According to its creators, the alignment with the FR family was crucial. As stated,

"A key element in the design of RDA is its alignment with the conceptual models for bibliographic and authority data developed by the International Federation of Library Associations and Institutions (IFLA): Functional Requirements for Bibliographic Records [and] Functional Requirements for Authority Data."[9]

This paper uses the FR family and the RDA as a starting point but detects some problems and inconsistencies between these models. It sustains the basic semantics from these standards but rejects their structural formalism because it finds that it is quite problematic and lacks effectiveness in expressing highly machine-processable data. The effective processability of the data will be discussed in detail in the section "The Impact of the Representation Scheme's Selection: RDF versus ER."

Among the FR family, the terminology is inconsistent and, as we pass from the FRBR to FRAD and FRSAD, even the perception angle of the general model undergoes change. In FRBR (the first in order), there is no notion of the *name* as an entity. FRAD introduces this perception (FRAD also adds *family* as a new entity) and FRSAD makes a step forward and introduces the concept of *nomen* instead of the concept of *name*. Hence, despite the fact that each of the members of the FR family of models has been represented in RDF,[10] there is no established consolidated edition yet that combines the different angles using a common model and terminology (vocabulary).[11] These representations (one for each model) are available at IFLA's website.[12]

On the other hand, in the context of RDA there may be more consistency regarding terminology, but, as is well established in the relevant literature, there are significant differences between the two models, i.e. the FR family and RDA.[13,14,15] Due to these differences, there are no URIs, not even in the RDA registry, in the examples of our study.[16]

Given the above, the terms appearing in the figures are a selection from the three texts of the FR family. Thus, *nomen* (from FRSAD) is used instead of *name* (from FRAD) as a more abstract notion, and the attribute—property in the context of RDF—"has string" (from FRAD) is used to assign a specific literal to a *nomen*. In figures 2–5 we have used the "has appellation" (reversed "is appellation of") relationship of FRAD.[17]

**Notes about Terminology and Graphs: How to Read the Figures**

In this paper two different sorts of figures appear. This covers the need to compare two different models and pinpoint the differences between them and the problems that arise from selecting the ER model to express FRBR. An explanation of the two major models follows in the next subsection.

The first figure type follows the diagrams of the Entity–relationship model and is used in figure 1. In this case:

- The rectangles represent entities.

- The oval shapes represent attributes.

- The diamond-shaped boxes represent relationships.

The second figure type has been created according to the RDF graphical representations and is used in figures 2–5. In these cases:

- The oval shapes represent nodes that are identified by a URI and they could serve as objects or subjects for further expansion of the network. In figures 3–5 all the names were derived from the FR entities.

- The line connectors between nodes represent the predicates (i.e., they are properties) and should also serve as URIs.

- The rectangle shapes represent literals consisting of lexical form. Language code could apply in these cases. With or without language codes, these are the end points and they could not be subject to new connections.

We follow the common modeling of the language in RDF in which the literal itself contains a language code, for example `"example"@en` in standard Turtle syntax, or `<rdfs:label xml:lang="en">` in RDFS XML coding. We must note that this kind of modeling is quite a simplistic way of language modeling because there is no mechanism to declare more information about language, such as multiple scripts, which could apply in the context of the same language.

**The Impact of the Representation Scheme's Selection: RDF versus ER**

Nowadays, all the information on library catalogs is created through and stored in computers. This technological infrastructure provides specific methods and dictates limitations for the catalog's data management. Hence, every model must take into consideration the basic rationale of the technological infrastructure that will curate and process the data. Depending on the syntax capabilities of the representation model, the expression of what we want to express becomes reasonably easy and accurate since "semantics is always going to have a close relationship with the field of syntax."[18] This establishes a vital relationship between what we want to do and how computers can do it.

In this section we emphasize the limitations of the Entity Relationship (ER) implementation, which FRBR proposes, and denote how syntax affects expressiveness and, accordingly, functionality. Finally, we demonstrate how the selection of one implementation or another (in our case ER vs. RDF) has serious implications, both for cataloging rules and for cataloging practice.

Why do we compare these two specific models? The ER model is the base that has been selected from IFLA as a "conceptual framework" [19] for the development of FRBR, while FRBR is the conceptual model upon which RDA has been founded. Subsequently, RDA is also affected by the choice of ER model. On the other hand, RDF is the current conceptualization for resource description in the web of data. So, what kind of problems and conflicts arise from the implementations of each of these models?

The basic rationale of ER comprises three fundamental elements. There are entities; entities have attributes; and there are relationships between entities. It is also possible to declare cardinality constraints upon which the FR family builds.

Then again, RDF implies quite a different model. "The core structure of the abstract syntax is a set of triples, each consisting of a subject, a predicate and an object. A set of such triples is called an RDF graph. An RDF graph can be visualized as a node and directed-arc diagram, in which each triple is represented as a node-arc-node link. . . . There can be three kinds of nodes in an RDF graph: IRIs, literals, and blank nodes."[20] "Linking the object of one statement to the subject of another, via URIs, results in a chain of linked statements, or linked data. This avoids the ambiguity of using natural language strings as headings to match statements. As a result, a literal object terminates a linked data chain, and literals are generally used for human-readable display data such as labels, notes, names, and so on."[21]

As a representative example of the differences between the two models, let us consider "place of publication." Peponakis counts nine attributes of place and notices that, due to the fact that the ER model does not allow links between attributes, there is no way to define explicitly whether these attributes address the same place or not.[22] Taking into consideration this problem we demonstrate the transition from the ER attributes approach to RDF implementations in figures 1–2.

Let us assume that there is Person (X), who was born in London, is named John Smith and works at Publisher (Y). This Publisher is located in London, where Book (1), entitled *History of London*, has been published. For this specific book, Person X was the lithographer. If we create a strict mapping to FRBR entities, attributes, and relations, then we have the situation illustrated in figure 1. Due to the fact that there is no way to link the four occurrences of London (inasmuch as there is no option to define relations between attributes in the ER model), there is no way to be certain that London is the same in all cases. Judging only by the name, it could stand for London in England, in Ontario, in Ohio, or elsewhere.

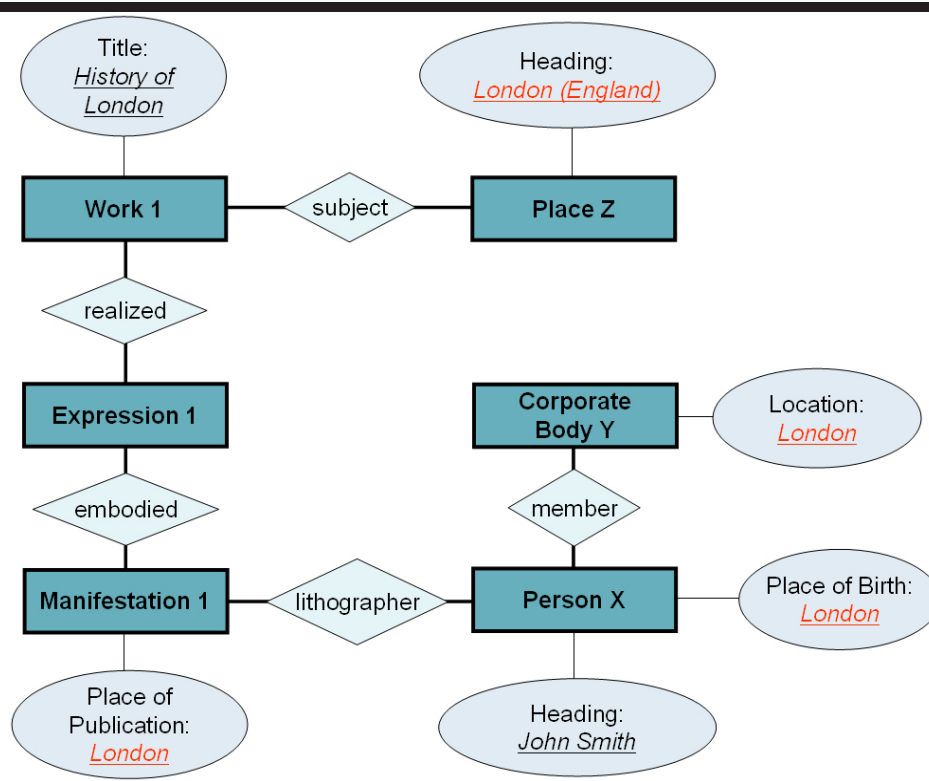

**Figure 1**. Example of "Place" as attribute of several entities

The IFLA working group has faced the problem with place and noted the following.

> *The model does not, however, parallel entity relationships with attributes in all cases where such parallels could be drawn. For example, "place of publication/distribution" is defined as an attribute of the manifestation to reflect the statement appearing in the manifestation itself that indicates where it was published. Inasmuch as the model also defines place as an entity it would have been possible to define an additional relationship linking the entity place either directly to the manifestation or indirectly through the entities person and corporate body which in turn are linked through the production relationship to the manifestation. To produce a fully developed data model further definition of that kind would be appropriate. But for the purposes of this study it was deemed unnecessary to have the conceptual model reflect all such possibilities.* [23]

Finally, they seem to avoid the problem and repeat their position in FRAD as well.

> *In certain instances, the model treats an association between one entity and another simply as an attribute of the first entity. For example, the association between a **person** and the **place** in which the person was born could be expressed logically by defining a relationship ("born in") between **person** and **place**. However, for the purposes of this study, it was deemed sufficient to treat place of birth simply as an attribute of **person**.* [24]

For some reason the creators of the FR family have chosen not to "upgrade" the attributes of place into one and only one entity. Furthermore, the same problem exists for many attributes, not only for place. Thus, the problem has to do with the selection of ER as "conceptual framework" and not with the specific entity of place. If we accept that "Place of Publication" must not be recorded as it appears on the resource, an RDF-based approach makes things clearer, as figure 2 shows. In this case, all attributes of place are promoted to the same RDF node and, instead of four repeats of the attribute with the value "London," we reduce it to one and only one node with four connections to it. Then, as illustrated by figure 2, we can be sure that all instances refer to the same London.

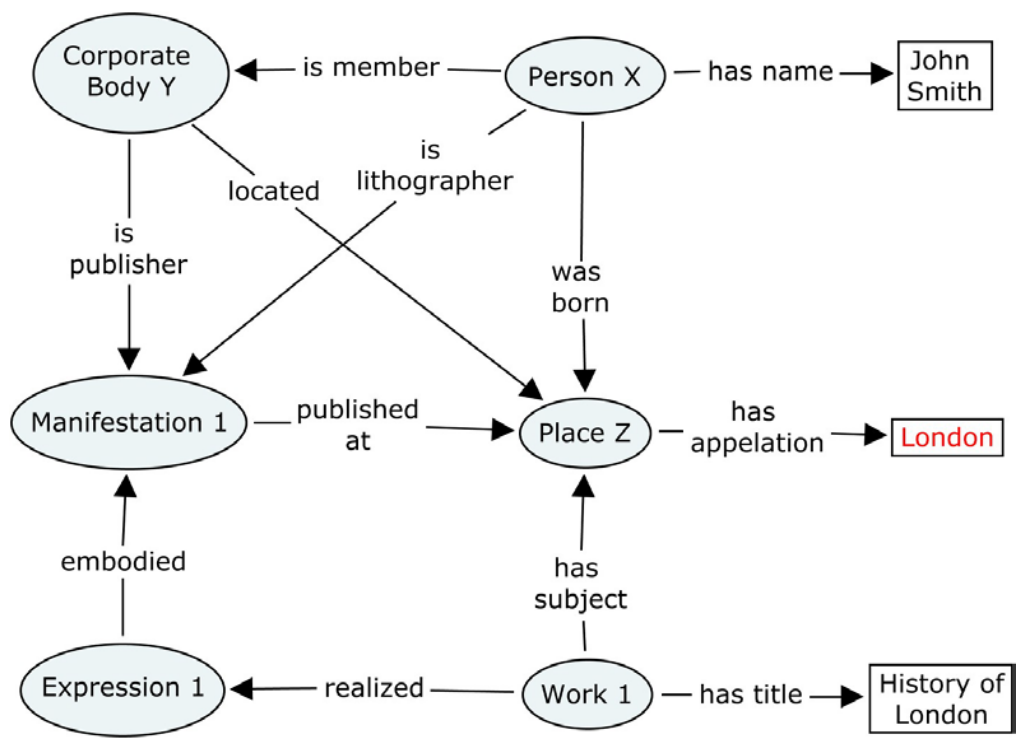

**Figure 2.** RDF-based representations of figure 1

In figure 2, it is assumed that there is no need to transcribe the literal of "Place of Publication" from the resource; i.e., we did not follow rule 2.8.1.4 of RDA: "Transcribe places of publication and publishers' names as they appear on the source of information." For cataloging rules that demand to record the place as it appears on the resource, the readers can consult the subsection "Place Names" in this study.

Last but not least, RDF has another significant advantage compared to the ER model: data coded in RDF are packed ready for use in the Semantic Web. On the contrary, data coded in ER must undergo conversion—with all its implications—in order to be published in the Semantic Web.

## NAMES, ENTITIES, and IDENTITIES

In this section, the significance of names as carriers of meaning is outlined and the importance of documenting the relations of names with the entities and identities they refer to is established. Additionally, the basic approaches are presented for metadata generation for managing names. These approaches resulted in the distinction (dissociation of authorities) from the bibliographic records, which in turn led (both FRBR/FRAD and RDA) to the lack of potentially linking—in an explicit way—the entity with the names it goes by. This linking, as it is presented later in this text, is fundamental for the description and interpretation of the entity.

In everyday communication, the usage of a name in a sentence plays the role of the identifier for the entity that this specific name indicates. If the speakers share a common background, there is no need for qualifiers other than the name in order to disambiguate information such as whether Nick is Person X or Person Y, or if the word "London" indicates the city in Ohio or in England, etc. Thus, the common background leads to a very limited context in which the interpretation of the name and the assignment to the appropriate entity is sufficient and accurate. However, the context of the Internet is extended into a variety of possibilities, so there is need of a more precise way to identify specific entities.

In this regard, a very essential issue is the distinction between the properties of the name and the properties of the entity that is represented by the specific name. The word "John" could be recognized as an English name, but we jump to a logical flaw if we assume that John knows English. A representative example of this kind of inference (syllogism) can be found in Rayside and Campbell.[25]

Statement:

> "Man is a species of animal.
>
> Socrates is a man.
>
> Therefore, Socrates is a species of animal.
>
> . . .
>
> 'Man' is a three-lettered word.
>
> Socrates is a man.
>
> Therefore, Socrates is a three-lettered word."

Therefore the authorities of a catalog should embody a two-level modeling of the information they represent. The first has to do with the entities and the second with the names of these entities. Consequently, there is the need to find a way to pass from names to the entities they indicate; and, from entities, to the various appellations that these entities have.

In catalogs, it is kind of vague whether the change of a name signifies a new identity. Niu states: "For example: the maiden name and the married name of an agent are normally not considered two separate identities, yet one pseudonym used for writing fiction and another pseudonym used for writing scientific works are often considered two different identities of an agent."[26] Then there can be one individual with many identities. But there can also be one identity which incorporates many individuals: for example, a shared pseudonym for a group of authors. To deal with these problems, FRAD introduces the notion of persona, rejecting at the same time the idea that a person is equal to an individual. FRAD defines a person as an "individual or a persona or identity established or adopted by an individual or group."[27] The question that arises here is when the persona must be conceived as a new identity. Yet, FRAD does not make a sufficient judgment; instead, they refer to cataloguing rules. "Under some cataloguing rules, for example, authors are uniformly viewed as real individuals, and consequently specific instances of the bibliographic entity person always correspond to individuals. Under other cataloguing rules, however, authors may be viewed in certain circumstances as establishing more than one bibliographic identity, and in that case a specific instance of the bibliographic entity person may correspond to a persona adopted by an individual rather than to the individual per se."[28] So there is no specific guidance if, for example, in the case of "religious relationship,"[29] there must be one identity created with two alternative names or two different identities. Rule 9.2.2.8 in RDA does not elaborate further.

Still, even with the problem of identities solved, the matter of appellations itself could be extremely complicated, and this is widely addressed in relevant literature.[30,31,32] The VIAF project confirms this with an extremely huge data set .[33] Assigning all appellations as attributes is an easy way to model the variants of a name, but it is very simplistic because it "does not allow these appellations to have attributes of their own and neither does it allow the establishing of relationships among the appellations. . . . FRAD makes a big step forward: all appellations are defined as entities in their own right, thus allowing full modeling."[34] Of course, FRAD's approach is not a novelty in the domain of LIS since library catalogs have been modeling names since the era of MARC. In UNIMARC Authorities,[35] the control subfield $5 contains a coded value to indicate the relations between the names with values such as "k = name before the marriage," "i = name in religion," "d = acronym," etc., and in MARC 21 there is the corresponding subfield $w.[36] FRAD puts these values on a more consistent and abstract level. FRAD also defines "Relationships between Persons, Families, Corporate Bodies, and Works" in section 5.3 and "Relationships between their Various Names" in section 5.4.[37]

## The Distinction between Authorities and Descriptive Information

Since the days of card catalogs and for as long as MARC and AACR have been used, bibliographic records have set their grounds on the dichotomy between descriptive information and control access points. The various types of headings stand for control access points. The terminus of headings was the alphabetical sorting. With the advent of computers, they were used as string identifiers to cluster and retrieve relevant bibliographic records. These bibliographic records had a body of descriptive information that was transcribed from the resource and remained unchanged. So the headings were the keys to the records and the records were surrogates for documents.

"The elements of a bibliographic record . . . were designed to be read and comprehended by human beings, not by machines"[38]; established headings are not an exception. One of their basic characteristics was the precondition that they were unique in the context of a specific catalog, thereby avoiding ambiguity. In every case of synonymy, qualifiers (such as date of birth or profession) were added to disambiguate, while the names also played the role of a unique identifier. From this process, an issue emerges: the information that appears on the document has changed and the controlled name may be completely different from the name on the resource. This means that the cataloger performs a transformation of the information, and this transformation carries two dangers. First, by changing the name, there is the possibility of assigning the entity behind the name to a wrong entity. Second, by disturbing the correspondence between the information on the resource and the information on the record of the resource, the record becomes a problematic surrogate of the resource. To surpass this obstacle, traditional catalogs split the information into two different areas: one with the established forms, i.e., the headings; and the second with the purely descriptive information, i.e., the information that must be transcribed from the resource. This is the reason why traditional library catalogs put much effort into transcribing information from resources and very detailed guidelines have been developed.

On the other hand, current approaches on metadata creation (such as Dublin Core) seem to underestimate the importance of descriptive information while concentrating on the established forms of names. But how can we be sure that different literals communicate the same meaning? Does this kind of simplification, perhaps, cause problems regarding the integrity of the information? The names are not just sequences of characters (i.e., strings), but they carry latent information. It is known that there are women who wrote using male names (for example Mary Ann Evans wrote as George Eliot) and men who wrote by using female names. There are also nicknames for groups (e.g., "Richard Henry" is a pseudonym for the collaborative works of Richard Butler and Henry Chance Newton), etc. Therefore, it is important not to ignore names and the forms in which they appear on the resources, but to model them in such a way that integration between authorities and descriptive information is feasible, and the names are efficiently machine-processable.

## INTEGRATING AUTHORITIES WITH DESCRIPTIVE INFORMATION

As we have already stated, traditional library catalogs are built on the dichotomy between description and access points. This analysis aims to bring descriptive information and authorities closer, i.e. to connect the access point of catalogs with the description of the resource. The basic principle of the model presented in this section is to promote each verbal (lexical) representation of a name to a nomen, whether this form of the name derives from a controlled vocabulary or not.

In the cases that this form appears in a specific vocabulary, appropriate properties could be used to indicate such a relation.

In this section, some representative examples are presented. It is important to note, once again, that every node and relation in the following figures could (and must, in the context of the Semantic Web) be identified by a URI, except for the values in rectangles, which are RDF simple literals and therefore cannot be the subjects of further expansion. Thus, the concatenation is the following: Every individual (instance of the relevant class) acquires a URI. Every individual is connected through the "has appellation" property (acquires URI) to a nomen (also acquires URI) and these nomens end up connected to a plain RDF literal, which is in natural language wording and cannot be subjected to further analysis.

**Place Names**

The problem of place as an attribute in FRBR and FRAD has also been analyzed in the Background Analysis of the current paper, specifically in the subsection "The Impact of the Representation Scheme's Selection: RDF versus ER." Here, a solution to this problem that is compatible with the FRBR/RDA solution is proposed. By promoting every nomen of a place to an RDF node, there is the option of referring to the entity of place as a whole or to a specific appellation of this entity. So, the relation (property in the context of RDF) between the subjects of a work could be indicated by connecting Work X with Place Z. On the other hand, according to rule 2.8.1.4 of RDA, the place of publication for the manifestation must be transcribed as it appears on the source of information. But following the connections presented in figure 3, it is easy to assume that this specific nomen corresponds to the same entity, i.e., to the same place.

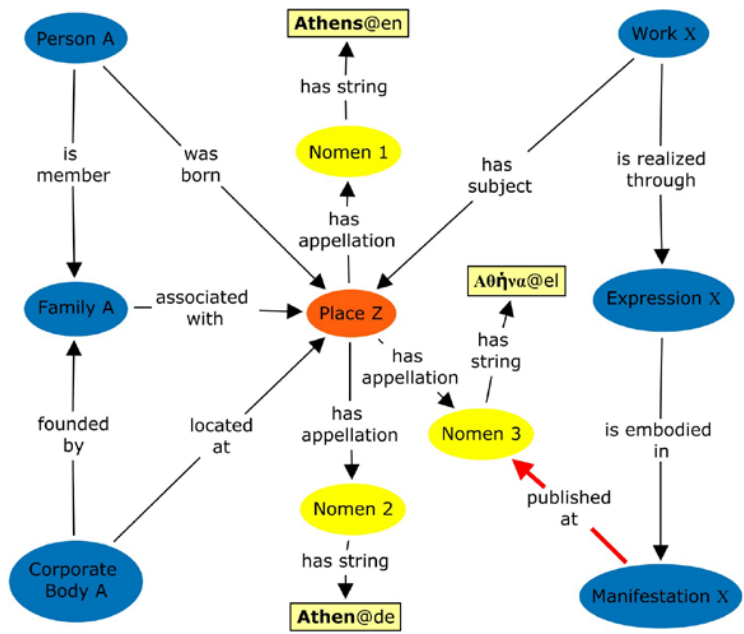

**Figure 3**. Place

**Personal names**

In the section "Names, Entities and Identities," we analyzed many of the problems associated with personal names. Here, a model is presented where the work (and expression) is connected directly with the author, whereas manifestation is connected with a specific appellation, i.e., nomen, of this author.

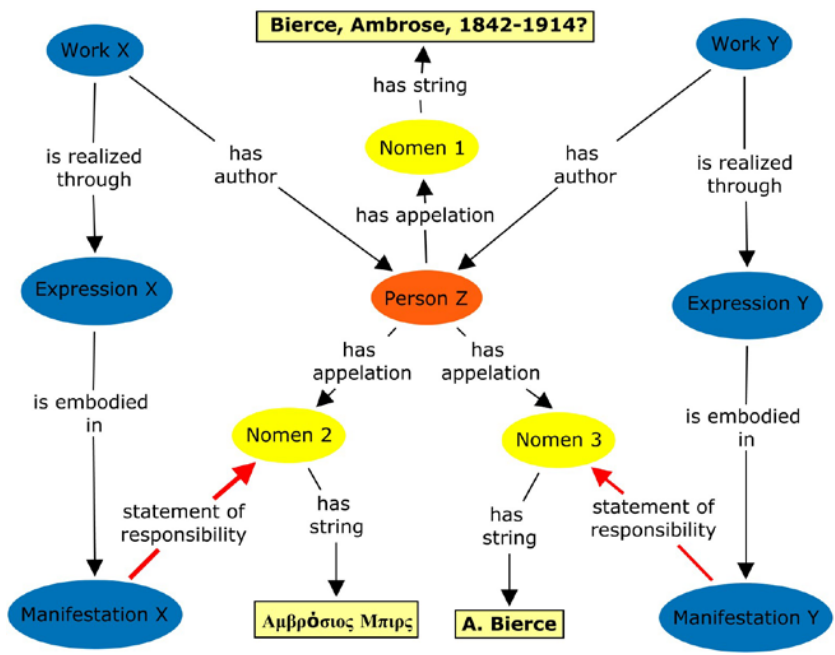

**Figure 4.** Statements of responsibility

RDA rule 2.4.1.4 states, "Transcribe a statement of responsibility as it appears on the source of information." But occasionally the statement of responsibility may contain phrases and not just names. In these cases, a solution similar to the Metadata Object Description Schema (MODS) could be implemented where, if needed, the statement of responsibility is included in the note element using the attribute type="statement Of Responsibility."

**Titles**

The management of titles in FRBR and RDA indicates a different point of view between the two standards. According to RDA there is no title for the expression,[39] and, as Taniguchi states, this is a "significant difference between FRBR and RDA."[40] BIBFRAME abides by the same principle of downgrading expression, since it entangles expression with work in an indivisible unit. In this regard, BIBFRAME is closer to RDA than to FRBR.

The notion of work has nothing to do with specific languages, even in the case when the work is a written text. Therefore the assignment of the title of work to a specific appellation is an unnecessary limitation. On the contrary, the title of a manifestation is derived by a specific resource. We argue that between these two poles there is the title of expression, which could stand as a uniform title per language.

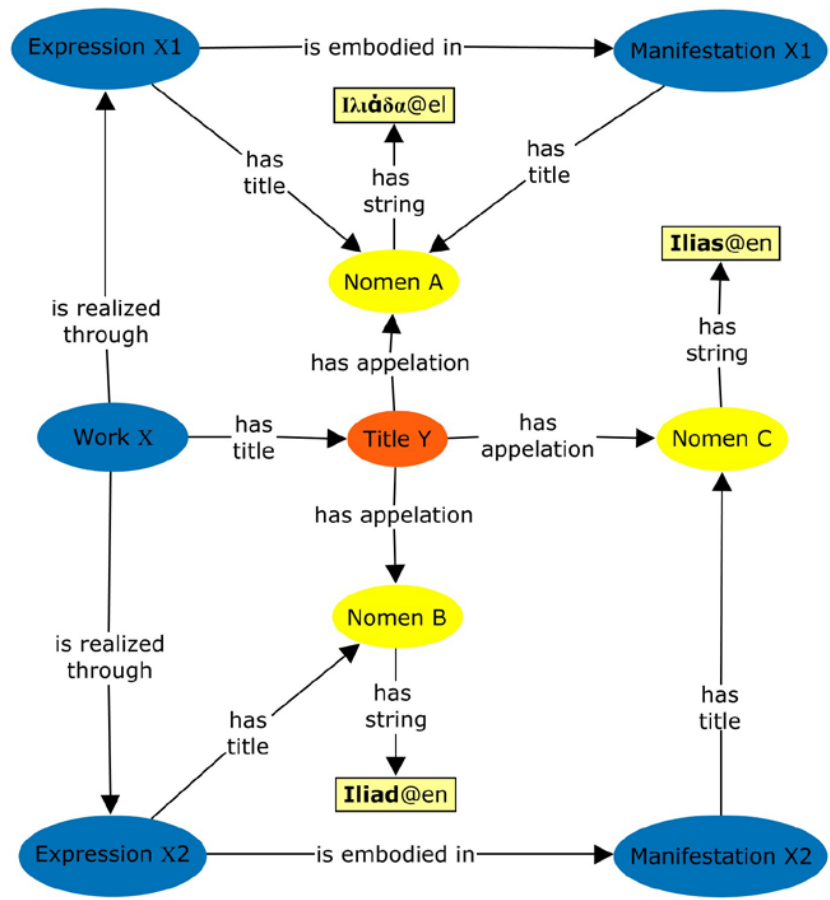

**Figure 5**. Titles

## VŁ) ˚žŁ˚( Ħ#" of BIBLIOGRAPHIC RECORDS and CATALOGING RULES

Resource description in the domain of LIS—from Cutter's era to the present day—emphasizes static linear textual representations. According to the RDA "0.1 Key Features," "In RDA, there is a clear line of separation between the guidelines and instructions on recording data and those on the presentation of data. This separation has been established in order to optimize flexibility in the storage and display of the data produced using RDA. Guidelines and instructions on recording data are covered in chapters 1 through 37; those on the presentation of data are covered in appendices D and E." But the tables in the relative appendices (D and E) contain guidelines that are mainly concentrated on punctuation issues, and they do not take into consideration the dynamics of current interactive user interface capabilities. As Coyle and Hillmann comment, "there are instructions for highly structured strings that are clearly not compatible with what we think of today as machine-manipulable data."[41] It is rather like producing high-tech cards: RDA is faithful to the classical text-centric approaches that produce bibliographic records as a linear enumeration of attributes; thus, RDA can be likened to a new suit that is quite old fashioned.

Traditional catalogs (from card catalogs to OPACs and repository catalogs) were built upon the principle of creating autonomous records. FRBR set this principle, i.e. one record for each resource, under dispute, while Linked Data abolishes it. This way, a gigantic graph of statements is created, while a certain part of these statements (not always the same) responds to or describes the desired information. Thus, a more sophisticated method emerges, if not makes itself imposed, for showing the results. Therefore, the issue is not to present a record that describes a specific resource, since this conceptualization tends to be obsolete altogether. Consequently, the visualization has to be different while in dependence with the data structure as well as the available interface of the searcher.

In this context, the analysis of this study tries to keep in balance the machine-processable character of RDF that builds on identifiers (URIs), while paying attention to the linguistic representation of entities. We argue that the balance between them will result in highly accurate and efficient representations for both humans and software agents. Let us consider the model for titles that has been introduced in this study. According to FRBR, "if the work has appeared under varying titles (differing in form, language, etc.), a bibliographic agency normally selects one of those titles as the basis of a 'uniform title' for purposes of consistency in naming and referencing the work."[42] RDA treats the case in a very similar way: rule 5.1.3 states, "The term 'title of the work' refers to a word, character, or group of words and/or characters by which a work is known. The term 'preferred title for the work' refers to the title or form of title chosen to identify the work. The preferred title is also the basis for the authorized access point representing that work". In this study, we consider the aforementioned statements as a projection that springs from the days when records were static textual descriptions independent of interfaces. Nowadays we are moving towards a much clearer distinction between the entity and its names. This is reflected in figure 5, in which the connection between a work and its author has nothing to do with specific names (appellations) but is based on URIs. The selection of the appropriate name as a title for the specific work could be based on certain criteria such as the language of the interface: in this case, the title of the work will be the title of the user interface language, and if this is not possible (i.e. there is no title label in this language), then it could be the title of the catalog's default language.

Following the kind of modeling proposed in the current study, the visualizations of data become more flexible and efficient in a variety of dynamic ways. Hence, we can isolate and display nodes and their connections, correlate them with the interface language or screen size (i.e., mobile phone or PC), create levels relative to the desired depth of analysis, personalize them upon the user's request or habits, and so on. Also, it becomes possible to display the data in forms other than textual. "As a result, humans, with their great visual pattern recognition skills, can comprehend data tremendously faster and more effectively through visualization than by reading the numerical or textual representation of the data."[43]

As we have already mentioned, the syntax and the semantics are always going to have a close relationship, but it is crystal clear that, now more than ever, the current Semantic Web standards allow for greater flexibility. As Dunsire et al. put it,

> *The RDF approach is very different from the traditional library catalog record exemplified by MARC21, where descriptions of multiple aspects of a resource are bound together by a specific syntax of tags, indicators, and subfields as a single identifiable stream of data that is manipulated as a whole. In RDF, the data must be separated out into single statements that can then be processed independently from one another; processing includes the aggregation of statements into a record-based view, but is not confined to any specific record schema or source for the data. Statements or triples can be mixed and matched from many different sources to form many different kinds of user-friendly displays.*[44]

In this framework, cataloging rules must reexamine their instructions in light of the new opportunities offered by technological advancements.

## DISCUSSION

Naming is a vital issue for human cultures. Names are not random sequences of characters or sounds that stand just as identifiers for the entities, but they also have socio-cultural meanings and interpretations. Recently, out of "political correctness" and fear of triggering racism, Sweden changed the names of bird species that could potentially offend, such as "gypsy bird" and "negro."[45] Therefore we cannot treat names just as random identifiers.

In this study we examined how, instead of describing indivisible resources, we could describe entities that appear in a variety of names on various resources. We proposed a method for connecting the names to the entities they represent and, at the same time, we documented the provenance of these names by connecting specific resources with specific names. We illustrated how to establish connections between entities, connections between an entity and a specific name of another entity, as well as connections between one name and another name concerning one or two entities. In the proposed framework, we maintain the linguistic character of naming while modeling the names in a machine-processable way. This formalism allows for a high level of expressiveness and flexible descriptions that do not have a static, text-centric orientation, since the central point is not the establishment of the text values (i.e., heading) but the meaning of our statements.

This study has shown that it is important to have the possibility to establish relationships both between entities and between specific appellations (nomens in the context of this study) of these entities. To achieve this we promoted every appellation to an RDF node. This is not something unheard of in the domain of RDF since this approach has also been adopted by W3C for the development of SKOS-XL.[46] FRBRoo, which is another interpretation of increasing influence in the wider context of the FR family, adopts the same perspective. [47] FRBRoo also gives the option to connect a specific name with a resource through the property "R64 used name (was name used by)" or to connect a name with someone who uses this specific name through the property "R63 named (was named by)."

Murray and Tillett state that "cataloging is a process of making observations on resources"[48]; hence, the production of records is the result of the judgments during this process. But in the context of traditional descriptive cataloging, the cataloger was not required to judge information in any way other than its category, i.e. to characterize whether the X set of characters corresponded to the name of an author, publisher, or place and so on. There was no obligation of assigning a particular name to a specific author, publisher, or place. In our approach, the cataloger interprets the information and supports the catalog's potential to deliver added-value information. Moreover, the initial information remains undifferentiated; hence, there is always the option of going back in order to generate new interpretations or validate existing ones.

In recent years, there has been a significant increase in the attention given to multi-entity models of resource description.[49] In this new environment, "the creation of one record per resource seems a deficient simplification."[50] RDF allows the transformation of universal bibliographic control to a giant global graph.[51] In this manner, current approaches on resource description "cannot be considered as simple metadata describing a specific resource but more like some kind of knowledge related to the resource."[52] Indeed, this knowledge can be computationally processable and exploitable. Yet, to achieve this, "catalogers can only begin to work in this way if they are not held bound by the traditional definitions and conceptualizations of bibliographic records."[53]

One critical issue is the isolation of parts (sets of statements) of this "giant graph" and the linking of these parts with something else; indeed, theory on this topic is starting to emerge.[54] This is very essential because it allows for the creation of ad hoc clusters (i.e. the usage of a specific identity for an entity with all the names that have been assigned to this identity, in our context), which could be used as a set to link to some other entity.

As a final remark, we could say that authorities manage controlled access points. In the Semantic Web, every URI is a controlled access point, and hence, the discrimination between description and authorities acquires a new meaning. In the context of machine-processable bibliographic data, the aim is to connect these two, i.e. the authorities with the description, and examine how one can support the other. However, since the emphasis is not on their individual management, we are drawn away from a mentality of 'descriptive information versus access points" and towards one of "descriptive information as an access point."

## ACKNOWLEDGEMENT

The author wishes to thank Henry Scott who assisted in the proofreading of the manuscript.